# Machining feature recognition using descriptors with range constraints for mechanical 3D models

Seungeun Lim[1,a], Changmo Yeo[1,b], Fazhi He[2,c], Jinwon Lee[3,d †], Duhwan Mun[1,e*]

[1] School of Mechanical Engineering, Korea University, Seoul 02841, Republic of Korea
[2] School of Computer Science, Wuhan University, Wuhan 430072, China
[3] Department of Industrial & Management Engineering, Gangneung-Wonju National University, Gangwon-do, 26403, Republic of Korea

[a]seungeunlim@korea.ac.kr, [b]ycm2420@korea.ac.kr, [c]fzhe@whu.edu.cn, [d]jwlee@gwnu.ac.kr, [e]dhmun@korea.ac.kr

†Co-corresponding author

* Corresponding Author, Tel: +82-2-3290-3359, Fax: +82-2-3290-5950

**Abstract**

In machining feature recognition, geometric elements generated in a three-dimensional computer-aided design model are identified. This technique is used in manufacturability evaluation, process planning, and tool path generation. Here, we propose a method of recognizing 16 types of machining features using descriptors, often used in shape-based part retrieval studies. The base face is selected for each feature type, and descriptors express the base face's *minimum, maximum,* and *equal* conditions. Furthermore, the similarity in the three conditions between the descriptors extracted from the target face and those from the base face is calculated. If the similarity is greater than or equal to the threshold, the target face is determined as the base face of the feature. Machining feature recognition tests were conducted for two test cases using the proposed method, and all machining features included in the test cases were successfully recognized. Also, it was confirmed through an additional test that the proposed method in this study showed better feature recognition performance than the latest artificial neural network.

# 1. Introduction

Machining features refer to specific shapes generated by cutting with machine tools in manufacturing parts. Typical forms include holes, pockets, slots, and fillets. Machining feature recognition means recognizing features from three-dimensional (3D) computer-aided design (CAD) models for parts. It is used in various applications, including manufacturability evaluation, process planning, and tool path generation.

Modern representative methods for recognizing machining features include graph-based, volume decomposition, and hint-based methods. Traditional algorithm-based methods are time-consuming owing to complex recognition algorithms. Furthermore, the recognition success rate for complex or intersecting features is significantly low.

In a previous study, we used descriptors to improve the shortcomings of conventional algorithm-based methods in shape-based recognition [1]. The descriptor-based method defines a base face for each feature. It expresses the properties of the base face, the relationship between the base face and adjacent faces, and adjacent faces in descriptors. Then, it recognizes features by calculating the similarity of descriptors for the feature's base face and a target face to be compared. The descriptor-based method can identify features fast because the similarity is calculated after quantifying each item comprising the descriptors. Also, since this method calculates the similarity by applying the concept of probability, it works robustly even when interference occurs between features.

However, the feature recognition method using descriptors proposed in the previous study had the following limitations. First, it misrecognized specific machining features such as closed pockets. Moreover, it could not recognize fillets and chamfers. Consequently, the feature recognition accuracy was low, and a rule-based recognition method had to be applied separately.

This study defines the improved descriptors with enhanced information expression by applying the concept of range constraint on the descriptors. In addition, we propose a feature recognition method based on enhanced descriptors. The descriptor developed in the previous study focused on the *minimum* constraint that the feature's base face must have. However, the improved descriptors consider the *maximum* and *equal* conditions of the feature's base face.

The main contributions of this study are as follows. We propose the concept of the descriptor using range constraints and establish machining feature recognition considering this. All target features in the two test cases could be recognized through similarity comparison using the improved descriptors. These characteristics increase the feature recognition rate compared to the previous descriptor-based method. Furthermore, the proposed method in this study showed better feature recognition performance than the latest artificial neural network in an additional test.

This paper is organized as follows. Section 2 reviews related works on feature recognition. Section 3 discusses the descriptors to which range constraints were applied. Section 4 presents the feature recognition method. Section 5 discusses the experimental results of recognized features for test cases. Finally, Section 6 presents the conclusions and prospects for future research.

## 2. Related Works

Recognition of machining features from 3D CAD models has been investigated thoroughly in the literature. Typical feature recognition methods include graph-based, volume decomposition, hint-based, similarity-based, Hybrid, and deep learning-based approaches.

The graph-based method recognizes features by analyzing whether the subgraph matches a specific feature pattern after expressing the relationship between the face and edge of the total shape as a graph structure. Joshi and Chang [2] recognized machining features by using an attributed adjacency graph (AAG) that encodes face-to-face adjacency relationships. In addition, a heuristic method for identifying the components of a graph was proposed. However, it had the problem of not recognizing the features of intersecting parts, such as T-slots. Chuang and Henderson [3] proposed a method that configures a shape graph (vertex-edge graph) from a solid model in the B-rep form, defines the regional shape patterns, and identifies patterns of machining features. Using the vertex-edge (V-E) graph is advantageous because it is easy to determine the patterns. However, since it only uses the V-E graph, it is limited to recognizing shape patterns as simply interconnected face shapes. Gavankar and Henderson [4] demonstrated that the protrusions and depressions in the edge-face graph of a B-rep model comprise biconnected components and proposed a method of separating such connected relationships. This graph theory has the advantage of the high efficiency of feature recognition, and it is easy to add new features to be recognized. However, it had problems such as inapplicability to blind holes and pockets that are open on two or more sides.

The volume decomposition method decomposes a volume with a complex shape into volumes with simple shapes and then recognizes machining features from those with simple shapes[5]. Volume decomposition methods are subdivided into detailed methods such as convex decomposition and cell-based decomposition.

The convex decomposition method generates volumes with simple shapes by decomposing a volume with a complex shape into the convex hull and delta volumes. Tang and Woo [6] attempted to recognize features using the alternating sum of volumes (ASV) technique, but it is disadvantageous in that decomposition does not converge for specific shapes. Kim [7] proposed alternating the sum of volumes with partitioning to address the drawbacks of conventional ASV and recognized features by recognizing unique volume shapes through this approach. The convex hull decomposition method can recognize features well, even for intersecting features that are not recognized by graph-based and hint-based methods. However, it has the disadvantage of not dealing with curved shapes such as rounds or fillets. The convex hull decomposition method mainly removes fillets or rounds in advance to solve this problem.

The cell-based decomposition method identifies features after decomposing shapes into simple cells and then composing a maximal volume by combining the cells. Kim and Mun [8] Proposed sequential and repetitive volume decomposition methods such as fillet-round-chamfer decomposition, wrap-around decomposition, volume split decomposition, and non-overlapping maximal volume decomposition. These methods have the advantage that the number of cells can be significantly reduced, and the volume can be decomposed faster than that in maximum volume decomposition. Sakurai and Dave [9] proposed a method of decomposing shapes into minimal cells with simple shapes by expanding the surfaces of objects and composing a maximal volume by combining such minimal cells. This method has the advantage that it can easily recognize many volumes. Woo [10] proposed a faster alternative to conventional cell-based decomposition methods. Here, the maximal volume of a solid input model is a large simple volume without any concave edges. This cell-based method is advantageous because it can be used even when features intersect, same as the convex hull decomposition method, and can recognize features even when a quadratic surface is included, unlike the convex hull decomposition method. However, the cell-based decomposition method has the disadvantage that it has the high time complexity to evaluate complex shapes.

The hint-based method assumes that random geometry or topology traces are left in the B-rep model. Therefore, the hint-based method recognizes features through geometric reasoning from the random geometry or topology trace, i.e., hints, instead of finding the complete pattern of features. The hint-based method has the limitations that it cannot recognize rules that have not been predefined, and individual recognition rules need to be defined for each feature. Vandenbrande and Requicha [11] developed an algorithm based on the volume intersection function that searches for hints from surfaces of predefined slots, holes, and pockets after decomposing the total volume into volumes that satisfy strict manufacturing conditions. Regli [12] developed an algorithm that explores hints using edge and vertex information rather than faces from a model. Han and Requicha [13] proposed a recognition algorithm for recognizing machining features such as slots and pockets. They developed the incremental feature finder that expanded the object-oriented feature finder(OOFF) features of Vandenbrande and Requicha [11]. Li et al. [14] proposed a hint-based approach to feature recognition for reusable shapes. This approach obtains generalized properties of the shape for generic feature recognition by using the shape variations or hints emerging during the modeling operations on vertices, edges, and faces. Then, generic and interacting features are recognized using generalized feature properties. Verma, A. and S. Rajotia [15] proposed a hint-based machining feature recognition system for 2.5D parts with arbitrary interacting features. This system used various algorithms to create hints for hole, linear slot, circular slot, floor-based pocket, and floorless pocket. Ranjan et al. [16] proposed a method to obtain the contour of a 2.5D machining part by projecting virtual ray on a virtual surface. They recognized the feature using the information of face and volume obtained from analysis results of boundary and length of rays. However, only orthogonal features such as plane and cylindrical surfaces are considered, and features such as counterbore holes and countersink holes are not.

The similarity-based method recognizes features based on predefined thresholds by inspecting the similarity between two shapes of a random feature $S_1$ and a predefined feature $S_2$. Hong et al. [17]first compared the overall shapes of machine parts and their detailed shapes by selecting suitable parts. Furthermore, they proposed simplifying parts through multi-resolution modeling, which adds or removes some parts. However, this method has the disadvantage that the overall shape cannot be compared accurately unless the parts can be appropriately simplified. Ohbuchi and Furuya [18] and Liu et al. [19] proposed a method comparing the similarity of shapes in images generated by a 3D model based on multi-view. Sánchez-Cruz and Bribiesca [20] proposed a method that measures the similarity by object transformation of a 3D model into a voxel form. However, it has the disadvantage that it is difficult to extract a series of features and essential elements for irregular objects such as motor vehicle parts. Yeo et al. [1] defined base faces for machining features and corresponding descriptors. They recognized features by measuring the similarity between descriptors of the faces comprising the input 3D CAD model of the B-rep form and predefined descriptors. Zehtaban et al. [21] proposed a framework to search for models similar to input CAD models. The framework uses the similarity retrieval module based on the characteristics obtained by the Opitz coding system. Opitz coding system is a method of technology applied in computer-aided manufacturing (CAM) for part classification, and includes information such as geometry, topology, dimensions, material, bore, forming, etc.

The hybrid method recognizes features using a combination of various feature recognition methods. Sunil et al. [22] proposed a hybrid (graph-based and rule-based) method to recognize interacting machining features from the B-rep model. In addition, they proposed a heuristic-hint based graph extraction method to easily recognize the n-side interacting feature without identifying the virtual links. Guo at al. [23] proposed another hybrid (graph-based and rule-based) method and weighted attribute adjacency matrix (WAAM) model to represent the data structure of the B-rep model.

Recently, 3D model classification and segmentation research using deep learning has been conducted to improve the shortcomings of conventional algorithms. Jian et al. [24] suggested an improved novel bat algorithm (NBA) combined with a graph-based method to utilize the backpropagation algorithm to supplement an existing neural network with a long training time. The improved NBA could extract composite features comprising similar features, which is impossible with the conventional AAG method. Zhang et al. [25] proposed a new network called PointwiseNet that combines low-level geometrics with high-level semantic information. PointwiseNet has the advantage that it is robust to input noises and has fewer end-to-end parameters. Lee et al. [26] suggested a 3D encoder-decoder network to regenerate 3D voxels, including machining features. Zhang et al. [27] presented a new framework that learns the functions of machining features using a 3D convolution neural network called FeatureNet, and it recognized 24 machining features accurately. Shi et al. [28] proposed a deep learning framework for feature recognition based on multiple sectional view (MSV) expressions called MSVNet. Peddireddy et al. [29] automatically synthesized feature data and synthetic mechanical parts to train a neural network. Furthermore, they proposed a method of identifying the machining process by combining a 3D convolution neural network and transfer learning. However, voxel-based expressions tend to lose information about fine parts of the model, and has the disadvantage of low accuracy for these parts. Yeo et al. [30] proposed a method of recognizing features by constructing a deep neural network with the descriptors extracted from machining features as input. The deep learning-based method has higher recognition speed and performance than the algorithm-based method, but it has the disadvantage that it requires a sufficient amount of training data beforehand.

Kim et al. [31] proposed a deep learning-based system to retrieve piping component catalog. This system recognizes the piping components using multi-view convolutional neural network(MVCNN) and PointNet after splitting the point clouds. MVCNN converts the focus of 3D objects to the focus of 2D images. PointNet classifies and subdivides the point clouds data. Colligan et al. [32] proposed a hierarchical B-rep graph to encode B-rep data. The hierarchical B-rep graph can represent the geometry and topology of the B-rep model and allows the B-rep model to be used as input to neural networks. In addition, they presented MFCAD++[33] which includes non-planar machining feature and more complex CAD model than the previous MFCAD. Shi et al. [34] proposed SSDNet which conducts feature segmentation and recognition using an object detection algorithm named single shot multibox detector(SSD). SSDNet takes a view image as input and predicts the types of all features and the 3D location in this view direction. Then the 3D bounding boxes achieved from different view directions are combined to form the result. Table 1 compares the proposed method to previous studies.

Traditional algorithm-based machining feature recognition research targets the B-rep models because many modern 3D CAD systems apply B-rep models to represent the shape. However, most artificial neural network-based studies, which have been reported a lot recently, use voxel models because voxel models have simple data structure and are easy to process with artificial neural networks. Recently, a few studies like [32] tried to use B-rep models for machining feature recognition.

Table 1. Comparison of the proposed method to previous studies

| Category | Related papers | Characteristics | Comparison to our method |
|---|---|---|---|
| Graph-based | Gavankar and Henderson [4] | - Failure to recognize blind holes and pockets<br>- Support of polyhedral solids | - Consideration of various holes such as counterbore and countersink holes<br>- Support of polyhedral and curved solids |
| Volume decomposition | Sakurai and Dave [9] | - Surfaces limited to plane and cylindrical surface | - Dealing with various surfaces such as plane, cylindrical, conical surfaces, etc. |
| | Woo [10] | - Increase in high-time complexity when dealing with complex shapes | - Fast recognition of features by the use of descriptors |
| Hint-based | Li et al. [14] | - Recognition of features after decomposing an original shape | - Recognition of features without decomposition |
| | Verma and Rajotia [15] | - Failure to recognize fillet and chamfer | - Recognition of fillet and chamfer |
| | Ranjan et al. [16] | - Surface limited to plane and cylindrical surface<br>- Exclusion of counterbore and countersink holes | - Consideration of countersink and counterbore holes |
| Similarity-based | Sánchez-Cruz and Bribiesca [20] | - Difficulty extracting characteristics and base elements from irregular features | - Possible to extract descriptors of each face in irregular features<br>- High accuracy and recall |
| | Zehtaban et al. [21] | - Low accuracy and recall | |
| Hybrid | Sunil et al. [22] | - Non-consideration of the island as a machining feature | - Consideration of the island as a machining feature |
| | Guo et al. [23] | - Lack of test cases such as complex and interacting features | - Use of test cases having complex and interacting features |
| Deep learning-based | Peddireddy et al. [29] | - Unable to process features small with respect to a 3D model | - Regardless of feature size with respect to a 3D model |
| | Colligan et al. [32] | - Difficulty in deciding the view direction when specific direction information is not available | - Recognition of machining features without additional information on the direction |
| | Shi et al. [34] | - Dealing with one machining feature existing on each face of a 3D model | - Dealing with multiple machining features existing on each face of a 3D model |

# 3. Descriptors for Machining Feature Recognition

## 3.1. Target Machining Features for Recognition

The following two assumptions were made to define the target machining features. First, the machining types were set to milling, drilling, and turning. Second, the machining method was set to 2.5 D. Under these two assumptions, 16 feature types were determined, as shown in Table 2: five features related to holes, two features related to slots, three features related to pockets, two features related to islands, two features related to fillets, and two features related to chamfers.

**Table 2. Machining features for recognition with descriptors**

| Type | Subtype | Descriptions |
|---|---|---|
| Holes | - Simple hole<br>- Counterbore hole<br>- Counterdrilled hole<br>- Countersink hole<br>- Taper hole | Circular enclosed volume removed from stock material. A hole may have a change in diameter represented by a taper or a step. |
| Slots | - Simple slot<br>- Floorless slot | Volume removed from stock material. This removal is represented by a sweeping square 'u' shape profile along a straight line. |
| Pockets | - Closed pocket<br>- Opened pocket<br>- Floorless pocket | Volume removed from stock material. This removal is represented with a planar bottom face and planar side faces. |
| Islands | - Closed island<br>- Opened island | Protruding volume contained inside the bottom face of a pocket. |
| Fillets | - Inner fillet<br>- Outer fillet | Volume removed along a chain of rounded edges from the stock material. |
| Chamfers | - Inner chamfer<br>- Outer chamfer | Volume removed along a chain of chamfered edges from the stock material. |

## 3.2. Determination of Base Faces for Machining Features

The base faces determined for the machining features discussed in the previous section are shown in Fig. 1. Base face refers to a face that can best represent the characteristics of machining features among multiple faces comprising the machining features. Also, face types associated with the base face are plane, conical surface, spherical surface, cylindrical surface, and toroidal surface. This study does not deal with the recognition of machining features in which the shape of the base face is a free-form surface in machining over 2.5D.

Here, the characteristics of machining features include machining type, machining shape, and machining parameters. In the case of holes, the base face varies depending on the combination of two or more different rotational shapes. For *simple holes* and *taper holes* comprising a single rotational shape, the hole's cylindrical face or conical face is selected as the base face. For *counterbore holes*, *countersink holes*, and *counterdrilled holes* comprising two or more different rotational shapes, a face other than the cylindrical face is selected as the base face. In the case of *slots* and *pockets*, the bottom face or side face is selected as the base face, depending on whether the bottom face exists. In the case of *islands*, the bottom face is selected as the base face. In the case of *fillets* and *chamfers*, the cylindrical face or planar face is selected as the base face.

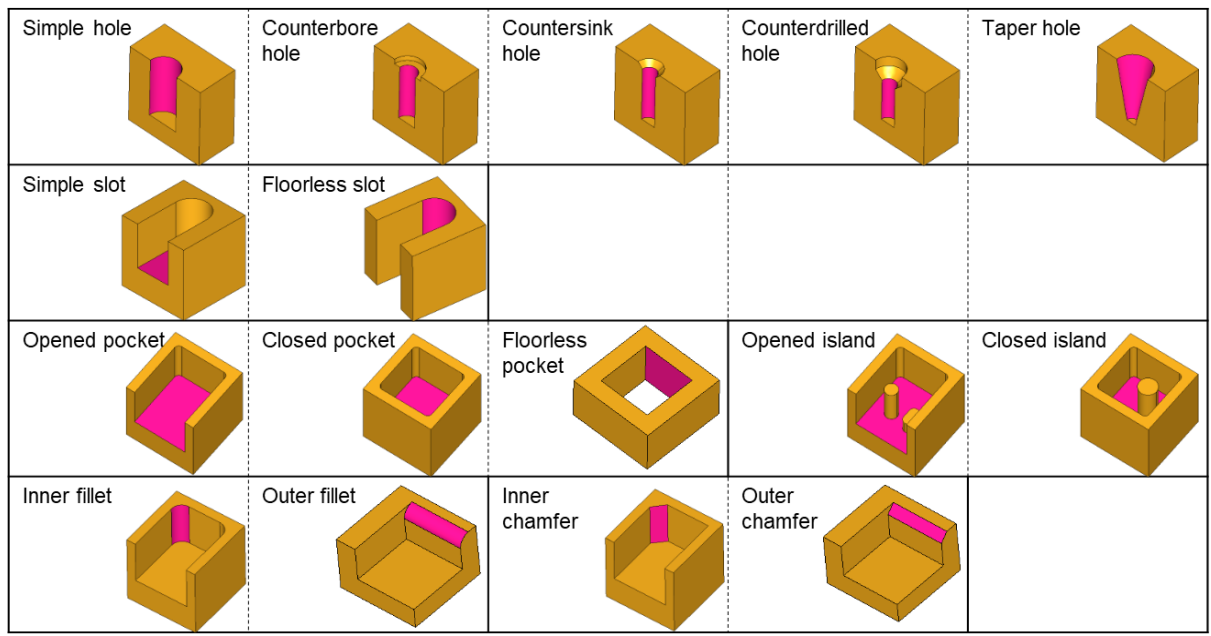

**Fig. 1. Base faces of machining features for recognition**

## 3.3 Definition of Descriptors
### 3.3.1 Concept of Descriptor
Each face comprising the B-rep model has the information about the face itself, edges comprising the face, vertices comprising the edges, and adjacent faces. The face information comprises the face type (e.g., planar face, cylindrical face, toroidal face, etc.), normal vector, and loop information (e.g., inner loop, outer loop). The edge information comprises the line type (e.g., linear type, curved type) and length, and the vertex contains the coordinate information. In addition, the relationships with adjacent faces can be identified. The relationship information comprises the angle between the target face and adjacent face, convexity, and continuity.

The descriptor is a data structure that expresses the information about the target face and the information about relationships to adjacent topology elements. As shown in Fig. 2, the information items expressed by the descriptor comprise the target face, outer loop, inner loop, and auxiliary information. The target face information comprises the *face type*, *curvature*, *face-machining*, *fillet-machining*, and *chamfer-machining*. The outer loop and inner loop information comprise the *convexity*, *continuity, parallel axis*, and *angle between the base face and adjacent face (acute angle, obtuse angle, and right angle)*. The auxiliary information comprises *parallel information between adjacent faces*, *coaxial matching information*, and *face interference information*.

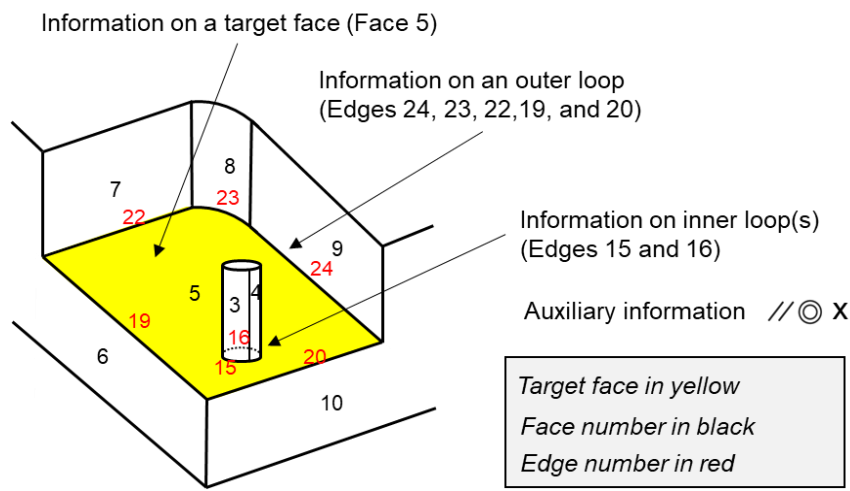

**Fig. 2. Concept of feature descriptor.**

The B-rep model can be represented with different geometric elements even if it has the same feature[35]. Especially, cylindrical shape is represented differently as one cylindrical face or two half-cylindrical faces according to the 3D CAD system. Therefore, we consider whether the descriptor can correspond to different expressions of the same shape when defining descriptors. If it is difficult to correspond to others, we define various descriptors to fit the feature representation. In this study, various descriptors of machining features in which base face is cylindrical are defined (e.g., descriptors of simple hole). Also, if the base face can be of various types, such as fillet or chamfer, various descriptors are defined.

### 3.3.2. Components of the descriptor

The details of the information comprising descriptors are as follows. The target face information comprises *face type item, curvature item, face-machining item, fillet-machining item, and chamfer-machining item*. The *face type item* is classified into planar, cylindrical, conical, spherical, and toroidal, expressed as PLAN, CYLI, CONI, SPHE, and TORO, respectively. If the face type cannot be specified, it is expressed as ANY. The *curvature item* is classified by the face's curvature (convex, flat, and concave). These are expressed as positive, flat, and negative, respectively. The *face-machining item* is for distinguishing slots and pockets. For the *face-machining item*, the pair of faces parallel to each other ($F_1$ and $F_2$) among the adjacent faces in contact with the outer loop of the corresponding face is considered, and the width of $F_1$ and $F_2$ is determined, which is then compared with the user input value. If the result is larger than the user input value. In that case, it is expressed as 'Longer.' If it is shorter, it is expressed as 'Shorter.' Moreover, if a face does not have a parallel face, it is expressed as 'Longer.' The *fillet-machining item* distinguishes fillets from other features. The *chamfer-machining item* distinguishes chamfers from other features. The shortest width for these two is obtained among the target faces and compared with the user input value. Suppose the width is longer than the user input value. In that case, it is expressed as 'Longer.' If it is shorter, it is expressed as 'Shorter.' If the type of the target face is a rotational, such as cylindrical, spherical, or toroidal, this value is used as the radius of the target face.

The loop information of the descriptor includes outer loop and inner loop information. The loop information comprises *convexity item*, *continuity item*, *parallel item*, *acute angle item, right angle item,* and *obtuse angle item*. The expression format of each item is as follows:

$$\text{face type } F_t \mid \text{convexity } C: \text{number of faces } N \qquad (1)$$

This is interpreted as "A target face has *N adjacent faces* of *face type Ft* and *convexity C* in the loop." The *convexity item* refers to the convexity between the target face and the adjacent face, which can be calculated as shown in Fig. 3. If $F_a$ and $F_b$ are the target face and adjacent face, respectively, the direction vector $\vec{d_r}$ is obtained by calculating the cross product of the normal vector of $F_a$ ($\vec{n_a}$) And the normal vector of $F_b$ ($\vec{n_b}$). And then, the dot product of $\vec{d_c}$ and $\vec{d_r}$ (the coedge of $F_a$ that is in contact with $F_b$) is calculated. If the calculated value is positivie, it is convex, and if the value is negative, it is concave. After calculating the convexity, the *convexity item* is expressed by adding the face type and the number of faces, as shown in Eq. (1). A *continuity item* implies continuity between the target face and the adjacent face. Here, continuity is distinguished between C0 and other continuity. Furthermore, the *continuity item* is expressed only for the adjacent face for which the continuity is not C0. The *parallel axis item* indicates whether the base vector $V_1$ of the target face is parallel to the base vector $V_2$ of the adjacent face. The base vector is an axial vector if the target face is a rotational face and a normal vector at the center of the face if it is not a rotational face. The *acute, perpendicular, and obtuse items* refer to the angles formed by the base vector $V_1$ of the target face and the base vector $V_2$ of the adjacent face. The expressions of these are as follows:

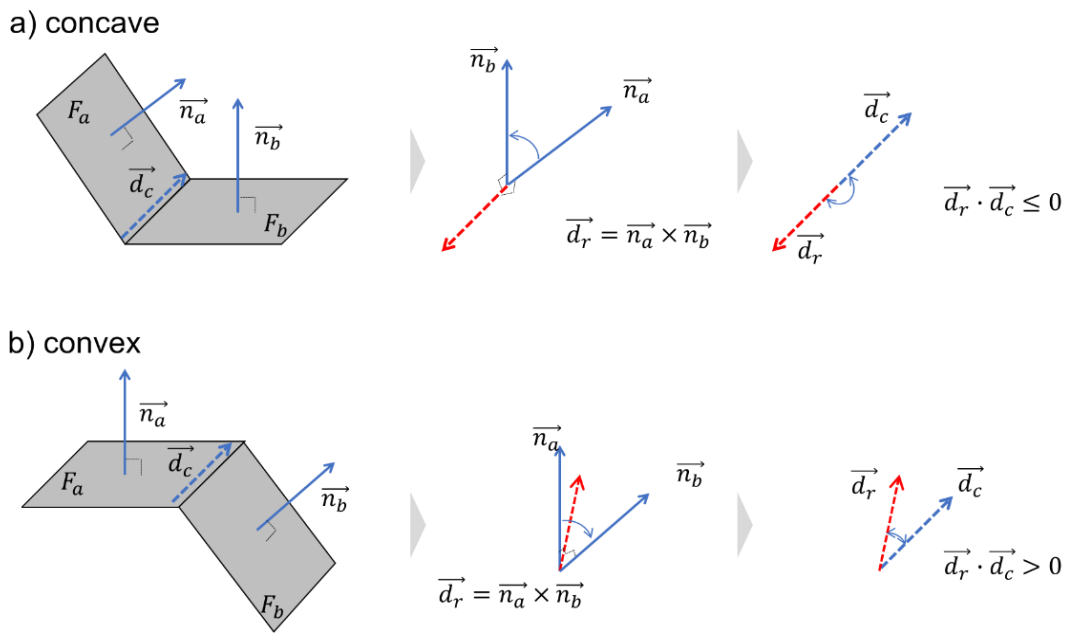

**Fig. 3. Calculation of convexity.**

Auxiliary information comprises *parallel item*, *coaxial item*, and *interference item*. The *parallel item* indicates whether the adjacent faces in contact with the outer loop of the target face are parallel to each other. If there is a pair of adjacent faces parallel to each other, it is expressed as 'True' or 'False.' Otherwise, it is expressed as 'False.' The *coaxial item* indicates whether the adjacent face F1 in contact with the outer loop of the target face and the adjacent face F2 in contact with the inner loop are rotational faces with the same axis. If this is the case, it is expressed as 'True,' and otherwise, it is expressed as 'False.' The *interference item* indicates whether there is another face in the normal direction of the target face. This information is obtained by investigating whether a ray projected in the normal direction intersects with another face. If they intersect, it is expressed as 'True,' and otherwise as 'False.' The above descriptor items are summarized in Table 3.

**Table 3. Information items constituting a descriptor**

| Descriptor item | | Descriptions | Expression |
|---|---|---|---|
| Face | $D_{f\_facetype}$ | The base face type | PLAN, CYLI, CONI, SPHE, or TORO |
| | $D_{f\_curvature}$ | The base face curvature | POSITIVE, NEGATIVE, or FLAT |
| | $D_{f\_facemachning}$ | Level of width between two parallel faces adjacent to the base face in the outer loop. It is used for distinguishing slots from other machining features. | LONGER or SHORTER |
| | $D_{f\_filletmachining}$ | Level of the width of the base face. It is used for distinguishing fillets from other machining features. | LONGER or SHORTER |
| | $D_{f\_chamfermachining}$ | Level of the width of the base face. It is used for distinguishing chamfers from other machining features. | LONGER or SHORTER |
| Outer loop | $D_{ol\_convexity}$ | Convexity between the base face and adjacent faces in the outer loop | face type | convexity: count |
| | $D_{ol\_continuity}$ | Continuity higher than c0 between the base face and adjacent faces in the outer loop | face type | convexity: count |
| | $D_{ol\_parallel}$ | Parallelism between the base vector ($v_1$) of the base face and the base vector ($v_2$) of the adjacent faces in the outer loop | face type | convexity: count |
| | $D_{ol\_perpendicular}$ | Perpendicularity between the base vector ($v_1$) of the base face and the base vector ($v_2$) of adjacent faces in the outer loop | face type | convexity: count |
| | $D_{ol\_acute}$ | Acute angle between the base vector ($v_1$) of the base face and the base vector ($v_2$) of adjacent faces in the outer loop | face type | convexity: count |

| | | | |
|---|---|---|---|
| Inner loops | $D_{ol\_obtuse}$ | Obtuse angle between the base vector ($v_1$) of the base face and the base vector ($v_2$) of adjacent faces in the outer loop | face type | convexity: count |
| | $D_{il\_convexity}$ | Convexity between the base face and adjacent faces in the inner loop | face type | convexity: count |
| | $D_{il\_continuity}$ | Continuity higher than c0 between the base face and adjacent faces in the inner loops | face type | convexity: count |
| | $D_{il\_parallel}$ | Parallelism between the base vector ($v_1$) of the base face and the base vector ($v_2$) of adjacent faces in the outer loop | face type | convexity: count |
| | $D_{il\_perpendicular}$ | Perpendicularity between the base vector ($v_1$) of the base face and the base vector ($v_2$) of adjacent faces in the inner loops | face type | convexity: count |
| | $D_{il\_acute}$ | Acute angle between the base vector ($v_1$) of the base face and the base vector ($v_2$) of adjacent faces in the inner loops | face type | convexity: count |
| | $D_{il\_obtuse}$ | Obtuse angle between the base vector ($v_1$) of the base face and the base vector ($v_2$) of adjacent faces in the inner loops | face type | convexity: count |
| Auxiliary | $D_{ax\_parallel}$ | Parallelism between adjacent faces in the outer loop | True or False |
| | $D_{ax\_coaxial}$ | Coaxiality between one adjacent face in the outer loop and one adjacent face in the inner loop | True or False |
| | $D_{ax\_interference}$ | Interference of the base face caused by other faces of a model | True or False |

### 3.3.3 Consideration of Range Constraints

When the feature's base face descriptor is compared with the target face descriptor, Each descriptor item is compared. Three cases can be obtained by examining the values for each descriptor item, as shown in Fig. 4. The first case is where the descriptor value must be between the lower and upper limits, as shown in $D_i$ in Fig. 4. The second case is where the descriptor must have a specific value, as shown in $D_j$ in Fig. 4. The third case is where the descriptor must have a value higher or lower than a specific value, as shown in $D_k$ in Fig. 4.

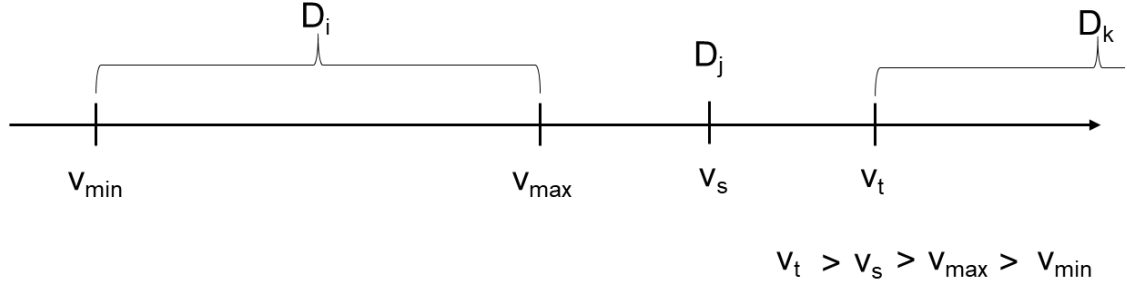

$$V_t > V_s > V_{max} > V_{min}$$

**Fig. 4. Possible ranges of value for each descriptor item**

Considering all the above three cases leads to describing the range constraints, i.e., *minimum*, *maximum*, and *equal* constraints, for each item of the descriptors for the feature base face. To this end, the descriptors defined in Table 3. are expanded, as shown in Fig. 5, to reflect the three range constraints for the feature base face. Here, the *minimum* and *maximum constraints* refer to the upper and lower bounds that each item of the descriptor can have. Furthermore, the *equal* constraint refers to the value that each descriptor item must have. The descriptor item for the target face and auxiliary has a value corresponding to the *equal* constraint. Furthermore, the descriptor item for the outer and inner loops has values corresponding to the *minimum* and *maximum* constraints. The method of recognizing the machining features by applying range constraints to the descriptor of the feature base face is explained in Section 4.

| Descriptor of a general face | |
|---|---|
| Descriptor item | Value |
| $D_{f\_facetype}$ | PLAN |
| $D_{f\_curvature}$ | FLAT |
| ⋮ | ⋮ |
| $D_{ol\_convexity}$ | CYLI \| CONCAVE : 2<br>PLAN \| CONVEX : 1 |
| ⋮ | ⋮ |
| $D_{il\_convexity}$ | CYLI \| CONVEX : 2 |
| $D_{il\_perpendicular}$ | CYLI \| CONVEX : 2 |
| ⋮ | ⋮ |
| $D_{ax\_coaxial}$ | TRUE |

Extension →

| Descriptor for the base face of a machining feature | | | |
|---|---|---|---|
| Descriptor item | Minimum | Maximum | Equal |
| $D_{f\_facetype}$ | - | - | PLAN |
| $D_{f\_curvature}$ | - | - | FLAT |
| ⋮ | ⋮ | ⋮ | ⋮ |
| $D_{ol\_convexity}$ | ANY \| CONCAVE : 1 | - | - |
| ⋮ | ⋮ | ⋮ | ⋮ |
| $D_{il\_convexity}$ | CYLI \| CONVEX : 1 | ANY \| CONCAVE : 0 | - |
| $D_{il\_perpendicular}$ | CYLI \| CONVEX : 1 | - | - |
| ⋮ | ⋮ | ⋮ | ⋮ |
| $D_{ax\_coaxial}$ | - | - | TRUE |

**Fig. 5. Extension of the descriptor to consider range constraints for the base face of a machining feature**

## 4. Machining Feature Recognition Method
### 4.1 Machining Feature Recognition Process

The machining feature recognition process using descriptors is shown in Fig. 6. First, the input B-rep model is analyzed to obtain the geometry and topology information of the faces comprising the model. Then, one of the subtypes of the machining features predefined in Fig. 2. is selected, and features are recognized using the descriptors of the base face of the corresponding features. One of the faces comprising the input model is selected, and the descriptor is generated from the selected face. The generated descriptor follows the format in Table 3. According to the user input information of machining conditions, LONGER is written in the descriptor item $D_{f\_facemachning}$, $D_{f\_filletmachining}$, and $D_{f\_chamfermachining}$ if the input value is larger than the width of the selected face (or distance between a pair of adjacent parallel faces); else, SHORTER is written. When descriptor generation from the selected face is completed, the similarity between the selected descriptor of the target face and the feature descriptor of the base face is calculated. If the similarity is greater than or equal to the threshold, the selected face is determined as the base face of the corresponding feature. The details of the similarity calculation between two descriptors are described in Section 4.2. Recognition for a specific feature is terminated when the similarity calculation for every face is completed. And then, other machining features defined in the list are recognized using the same method. Finally, the machining feature recognition result is output.

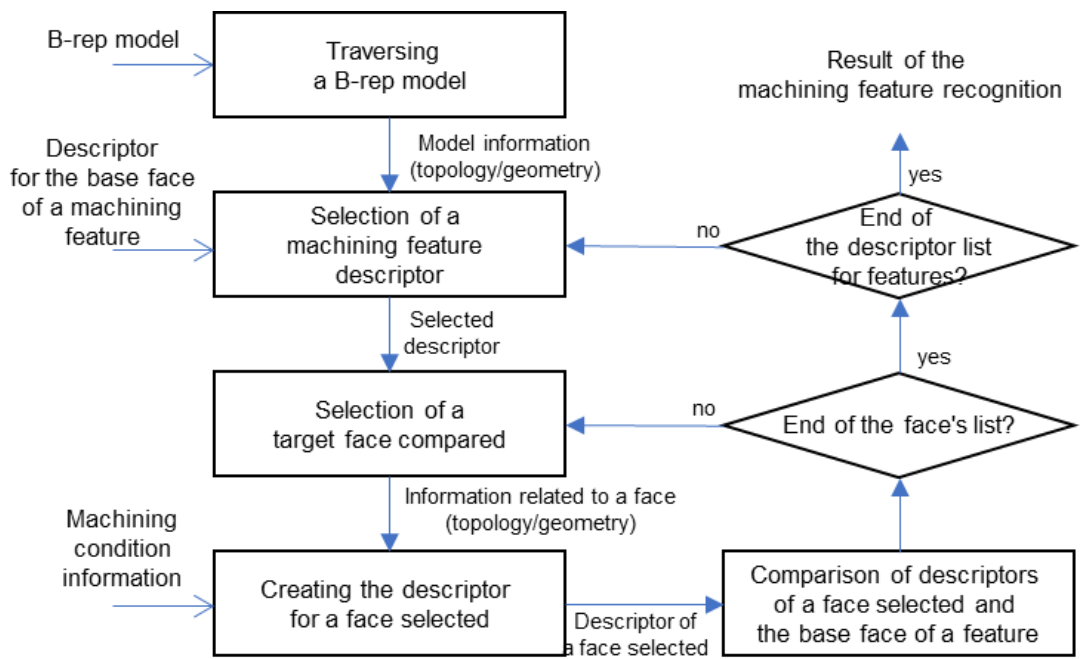

**Fig. 6. Procedure of machining feature recognition**

## 4.2 Feature Recognition using Descriptors based on Range Constraints

Fig. 7 shows the procedure for comparing the similarity of each descriptor. First, a descriptor item value is extracted from a target face after the descriptor item to compare the similarity to is selected. Then, similarity values ($A_{k,min}$, $A_{k,max}$, $A_{k,equal}$) for three range constraints *minimum, maximum,* and *equal* are calculated. When the similarity values are calculated for each range constraint, they are multiplied to obtain the similarity value $S_k$ for each descriptor item. This process is repeated for all descriptor items. When the similarity value for each descriptor item is calculated, the similarity value for each item is multiplied by the weight and added to obtain the similarity value of the descriptor(R). Finally, the similarity value of the descriptor is compared with the threshold value to determine whether the target face corresponds to the descriptor of a specific feature.

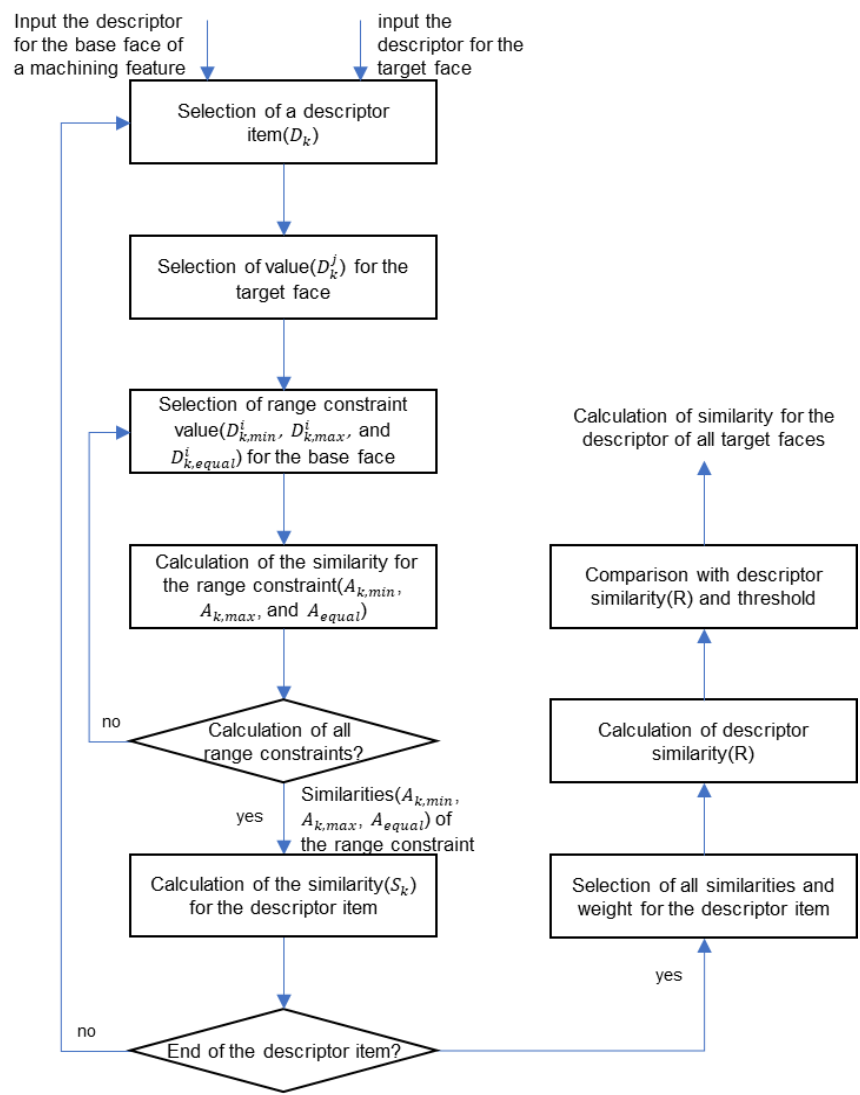

**Fig. 7. Procedure of similarity comparison for each descriptor**

It is assumed that the feature base face is $F^i$, target face to be compared is $F^j$, and values of the descriptor item $D_k$ of each face are denoted by $D_k^i$ and $D_k^j$, where $k$ is the descriptor item name described in Section 3.3.2. Furthermore, the descriptor item $D_k^i$ of the base face is subdivided again to $D_{k,min}^i$, $D_{k,max}^i$, and $D_{k,equal}^i$ according to the three constraints of *minimum, maximum,* and *equal*.

To calculate the similarity of the two faces ($F^i$, $F^j$), first, the similarities according to the *minimum, maximum,* and *equal* constraints ($A_{k,min}$, $A_{k,max}$, $A_{k,equal}$) is calculated for each descriptor item, as shown in Table 4. Then, as shown in Eq. (2), the similarity $S_k$ for each descriptor item is calculated by multiplying $A_{k,min}$, $A_{k,max}$, and $A_{k,equal}$.

**Table 4. Similarity comparison considering range constraints for each similarity item D$_k$**

| Type | Definition | Mathematical expression |
|---|---|---|
| Minimum | If $D_k^j$ is more than or the same as $D_k^i$ The result is one. If not, the result is zero. | $A_{k,min} = \begin{cases} 1 & (D_{k,min}^i \leq D_k^j) \\ 0 & (D_{k,min}^i > D_k^j) \\ 1 & (D_{k,min}^i \text{ is not specified}) \end{cases}$ |
| Maximum | If $D_k^j$ is less than or the same as $D_k^i$ The result is one. If not, the result is zero. | $A_{k,max} = \begin{cases} 1 & (D_{k,max}^i \geq D_k^j) \\ 0 & (D_{k,max}^i < D_k^j) \\ 1 & (D_{k,max}^i \text{ is not specified}) \end{cases}$ |
| Equal | If $D_k^j$ is the same as $D_k^i$ The result is one. If not, the result is zero. | $A_{k,equal} = \begin{cases} 1 & (D_{k,equal}^i = D_k^j) \\ 0 & (D_{k,equal}^i \neq D_k^j) \\ 1 & (D_{k,equal}^i \text{ is not specified}) \end{cases}$ |

$$S_k = A_{k,min} * A_{k,max} * A_{k,equal} \qquad (2)$$

When the similarity $S_k$ for the descriptor item $D_k$ is calculated, the final similarity R for two faces ($F^i$, $F^j$) is calculated as in Eq. (3). The weight of each descriptor item is determined to calculate the final similarity R. The sum of all weights is 1, as shown in Eq. (4). The default value of each weight is the same.

$$R = \sum_{S_k \in I, W_k \in J} S_k * W_k \qquad (3)$$

$$\sum_{W_k \in J} W_k = 1 \qquad (4)$$

where,

$$I = \begin{Bmatrix} W_{f\_facetype}, W_{f\_curvature}, W_{f\_facemachining}, W_{f\_filletmachining}, W_{f\_chamfermachining}, \\ W_{ol\_convexity}, W_{ol\_continuity}, W_{ol\_parallel}, W_{ol\_perpendicular}, W_{ol\_acute}, W_{ol\_obtuse}, \\ W_{il\_convexity}, W_{il\_continuity}, W_{il\_parallel}, W_{il\_perpendicular}, W_{il\_acute}, W_{il\_obtuse}, \\ W_{ax\_parallel}, W_{ax\_coaxial}, W_{ax\_interference} \end{Bmatrix},$$

$$J = \begin{Bmatrix} S_{f\_facetype}, S_{f\_curvature}, S_{f\_facemachining}, S_{f\_filletmachining}, S_{f\_chamfermachining}, \\ S_{ol\_convexity}, S_{ol\_continuity}, S_{ol\_parallel}, S_{ol\_perpendicular}, S_{ol\_acute}, S_{ol\_obtuse}, \\ S_{il\_convexity}, S_{il\_continuity}, S_{il\_parallel}, S_{il\_perpendicular}, S_{il\_acute}, S_{il\_obtuse}, \\ S_{ax\_parallel}, S_{ax\_coaxial}, S_{ax\_interference} \end{Bmatrix}$$

It is necessary to determine the magnitude relationship between the descriptor item values when calculating the similarity for each descriptor item. The calculation method is as follows. When feature base face $F^i$ and target face $F^j$ are compared, the descriptor item values are expressed as $F_t^i | C^i : N^i$, $F_t^j | C^j : N^j$, respectively. In this case, the magnitude relationship of descriptor items is calculated as follows. First, when face type and convexity are the same, the magnitude relationship between two descriptor item values is determined by the number of faces included in each descriptor item value. In other words, when the face type ($F_t$) and convexity (C) of two descriptor item values are the same, the number of faces $N^i$ and $N^j$ are compared. The method of comparison is shown in Eq. (5). Second, when face type ($F_t^i$) of the descriptor item value of base face is ANY, the number of faces $N^j$ of the target face's descriptor item value is the total number of faces ($N_k^j$) of the descriptor item value of the target face having the same convexity ($C_k^j$) as the convexity ($C^i$) of the corresponding descriptor item value.

$$\begin{aligned} F_t^i | C^i : N^i \text{ is less than } F_t^j | C^j : N^j, \text{ if } N^i < N^j, \\ F_t^i | C^i : N^i \text{ is greater than } F_t^j | C^j : N^j, \text{ if } N^i > N^j \\ F_t^i | C^i : N^i \text{ equals to } F_t^j | C^j : N^j, \text{ if } N^i = N^j \end{aligned} \qquad (5)$$

where $F_t^i = F_t^j$ and $C^i = C^j$ for two descriptor item values $F_t^i | C^i : N^i$ and $F_t^j | C^j : N^j$

$$N^j = \sum N_k^j \; for \; all \; fact \; type \; k \tag{6}$$

where $F_t^i = ANY$ and $C_k^j = C^i$ for two descriptor item values $F_t^i|C^i:N^i$ and $F_k^j|C_k^j:N_k^j$

If the final similarity R is greater than or equal to the predefined threshold, the target face $F^j$ is considered to be the base face of a specific machining feature. If there is no value of a specific descriptor item in the descriptor of the base face for a machining feature, the descriptor item is excluded from the similarity calculation. As a result, Eq. (5) is adjusted so that the sum of weights of the remaining items after excluding the corresponding item becomes 1.

Fig. 8 shows the process of comparing the descriptor $D^j$ of the target face to be compared with the descriptor $D^i$ of base face of the counterbore hole to identify the base face of the counterbore hole included in the 3D CAD model. The descriptor of the counterbore hole has only 3, 1, and 3 descriptor items for the *minimum*, *maximum*, and *equal* constraint, respectively. Therefore, only seven descriptor items are used in the similarity calculation, as explained above.

In the descriptor items of the *face* information, the $D_{f\_facetype,equal}^i$ of the feature's base face $F^i$ is PLAN in Fig. 8(a), and the $D_{f\_facetype,equal}^j$ of the target face $F^j$ is PLAN in Fig. 8(b). Because the two descriptors have the same value, the similarity $A_{f\_facetype,equal}$ becomes 1, according to Table 4. However, no value is given for $D_{f\_facetype,min}^i$ and $D_{f\_facetype,max}^i$. Therefore, $A_{f\_facetype,min}$ and $A_{f\_facetype,max}$ become 1, according to Table 4. Finally, the similarity $S_{f\_facetype}$ of the descriptor item $D_{f\_facetype}$ becomes 1 according to Eq. (2). In the same manner, calculating the descriptor item $D_{f\_curvature}$, the similarity $S_{f\_curvature}$ becomes 1

In the descriptor items of the *loop* information, $D_{ol\_convexity,min}^i$ of base face $F^i$ is ANY | CONCAVE : 1 in Fig. 8(a). Also, $D_{ol\_convexity}^j$ of the target face $F^j$ is CYLI | CONCAVE : 2 and PLAN | CONVEX: 1 in Fig. 8(b). According to the comparison method of the magnitude relationship of descriptor items described above, it becomes $D_{ol\_convexity,min}^i < D_{ol\_convexity}^j$. Therefore, the similarity $A_{ol\_convexity,min}$ becomes 1, according to Table 4. However, no value is given for $D_{ol\_convexity,max}^i$ and $D_{ol\_convexity,equal}^i$. Therefore, $A_{ol\_convexity,max}$

and $A_{ol\_convexity,equal}$ are 1, according to Table 4. Finally, the similarity $S_{ol\_convexity}$ of the descriptor item $D_{ol\_convexity}$ becomes 1, according to Eq. (2). In the same manner, calculating the descriptor items $D_{il\_convexity}$ and $D_{il\_perpendicular}$, the similarities $S_{il\_convexity}$ and $S_{il\_perpendicular}$ become 1

Finally, in the descriptor items of auxiliary information, the $D_{ax\_coaxial,equal}^{i}$ of the feature base face $F^i$ is TRUE in Fig. 8(a). Furthermore, the $D_{ax\_coaxial}^{j}$ of the target face $F^j$ is TRUE, in Fig. 8(b). Since these two descriptors have the same value, the similarity $A_{f\_facetype,equal}$ becomes 1, according to Table 4. However, no value is assigned to $D_{ax\_coaxial,min}^{i}$ and $D_{ax\_coaxial,max}^{i}$. Therefore, $A_{ax\_coaxial,min}$ and $A_{ax\_coaxial,max}$ are 1, according to Table 4. Finally, The similarity $S_{ax\_coaxial}$ of descriptor item $D_{ax\_coaxial}$ becomes 1, according to Eq. (2). When the calculation of similarity for every constraint is completed, as shown in Fig. 8(c), the similarity $S_k$ for each descriptor item is calculated in Fig. 8(d). Then, the final similarity R is calculated by reflecting a weight in the similarity $S_k$ for each descriptor item.

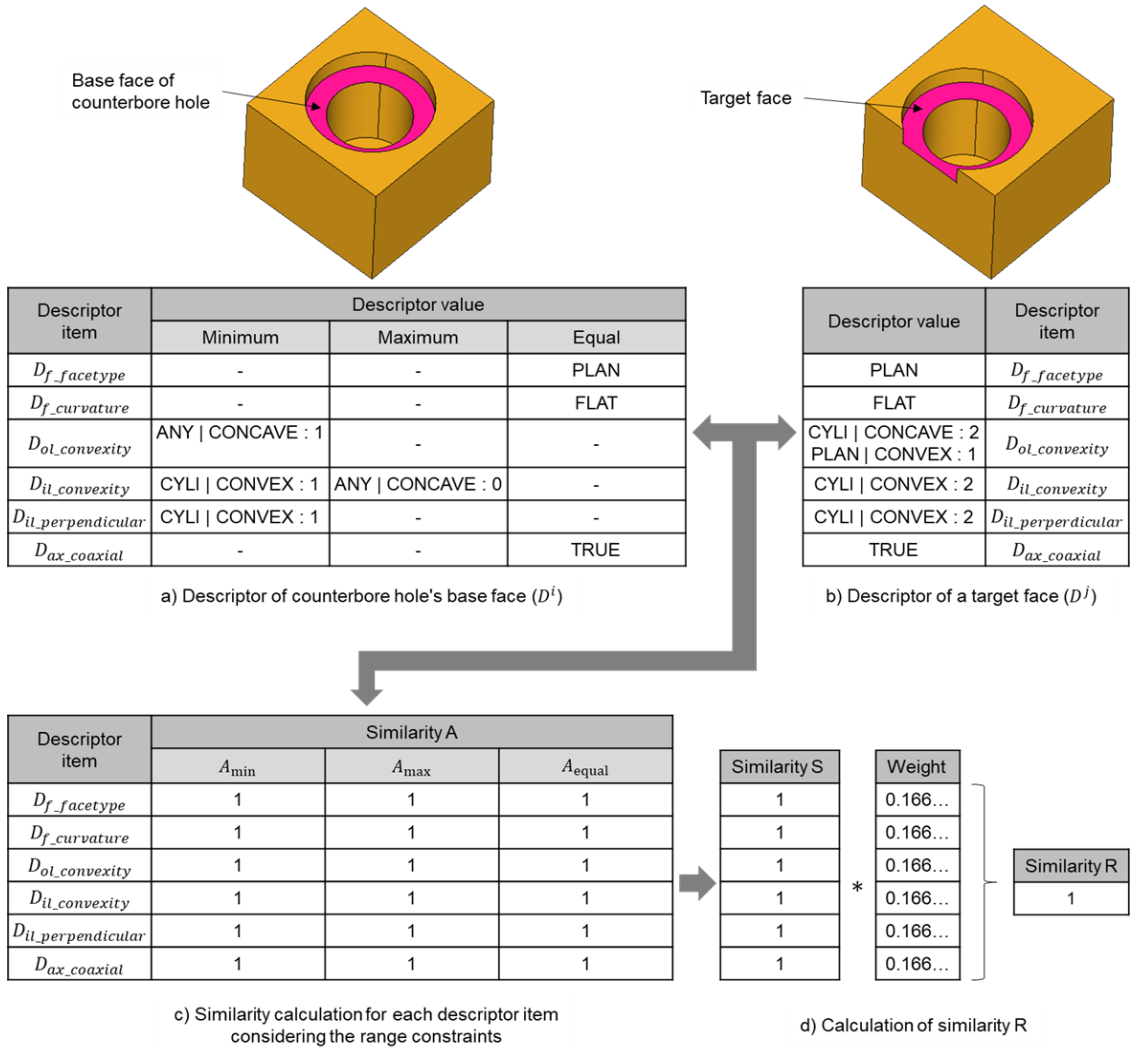

**Fig. 8. Similarity comparison between a target face and the descriptor of the counterbore hole's base face**

### 4.3. Processing Composite Features Described as Combination of Multiple Features

The concept of priority needs to be applied to the recognition process described above to correctly recognize composite features expressed as a set of multiple features. This study has three composite feature types: counterbore hole, counterdrilled hole, and countersink hole. As shown in Fig. 9, the countersink hole is a composite feature expressed as a combination of the simple and taper holes. If the feature is recognized by comparing the descriptor-based similarity, the simple hole, taper hole, and countersink hole are recognized simultaneously. To solve this problem, we prioritized that composite features are recognized before other general features.

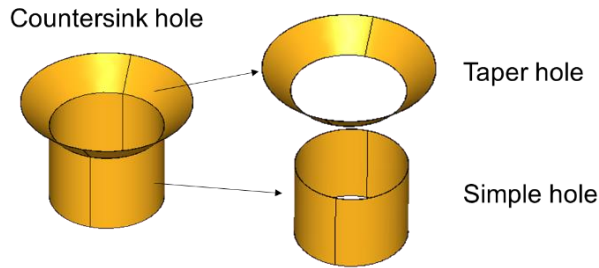

**Fig. 9. Composite feature represented by a combination of other features**

## 5. Implementation and Result

### 5.1. Machining Feature Recognition System and Recognition Performance Evaluation

As shown in Fig. 10, a prototype system was implemented to verify the recognition performance of machining features using the proposed descriptors. The prototype system was developed using the C++ programming language. Visual Studio 2019 was used for the integrated development environment, and the graphical user interface (GUI) was implemented using the Microsoft foundation class library. Open CASCADE 7.3.0 [36] was used for the shape modeling kernel.

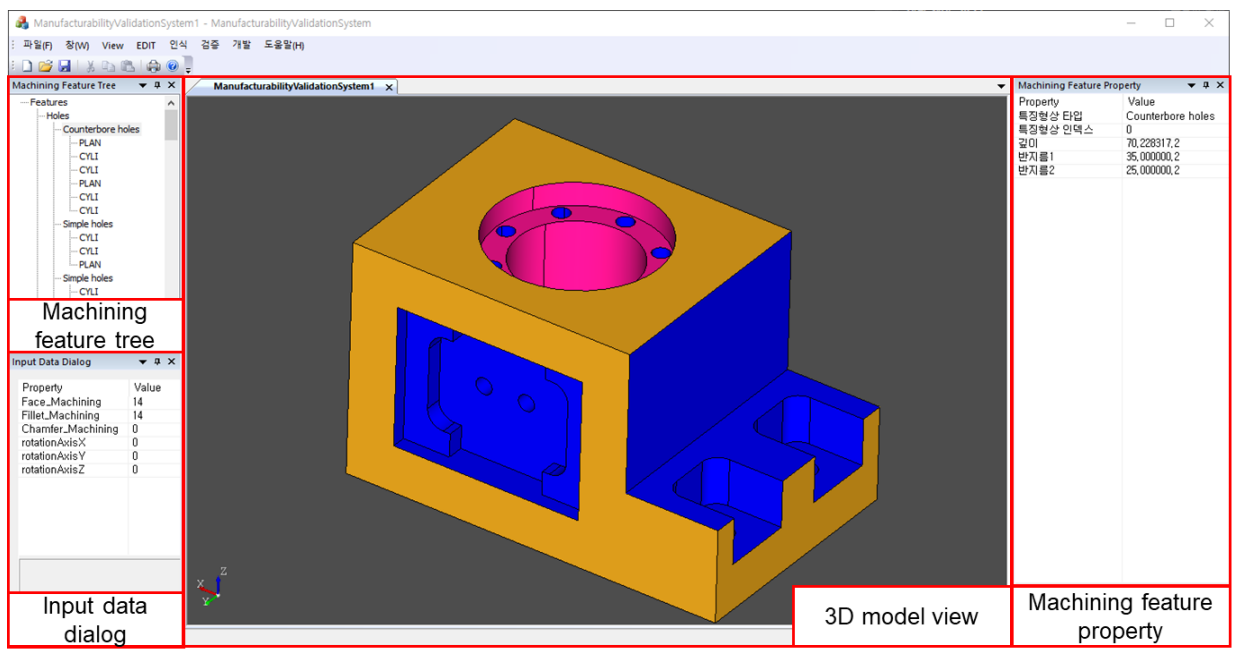

**Fig. 10. Prototype system used in the experiments**

When the recognition of machining features was tested using the implemented prototype system, the weight values for similarity comparison were evenly divided. In addition, the threshold for similarity R used to determine machining features was set to 1.

The test cases used in the machining feature recognition test are summarized in Fig. 11. Test case 1 was created by referring to the 3D CAD model used in the DFMPro solution of H company [37]. Furthermore, test case 2 was created by referring to the 3D CAD models used in related studies on feature recognition [38-42]. All machining features were successfully recognized in the two test cases using the proposed descriptor-based recognition method. In contrast, the previous study failed to recognize specific machining features, i.e., chamfer, simple hole, closed pocket, and floorless pocket included in the model nos. 8 and 12 of test case 1 and models nos. 1, 7, 10, and 12 of test case 2.

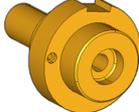

| No | List of machining feature | No | List of machining feature | No | List of machining feature |
|---|---|---|---|---|---|
| 1 | counterbore hole, simple slot, fillet, chamfer | 12 | simple hole, counterbore hole, simple slot, closed pocket, opened pocket, fillet | 20 | simple hole, counterbore hole, simple slot, closed pocket, opened pocket, fillet |
| 2 | counterbore hole, chamfer | | | | |
| 3 | counterbore hole, chamfer | 13 | simple hole, counterbore hole, simple slot, closed pocket, opened pocket, fillet | 21 | simple hole, counterbore hole, simple slot, closed pocket, opened pocket, fillet |
| 4 | counterbore hole, fillet, chamfer | | | | |
| 5 | counterbore hole | | | | |
| 6 | chamfer | 14 | simple slot | 22 | opened pocket |
| 7 | closed pocket, fillet | 15 | simple slot | 23 | opened pocket, fillet |
| 8 | closed pocket, chamfer | 16 | simple hole | 24 | simple hole, opened pocket |
| 9 | simple slot | 17 | simple hole | 25 | simple hole |
| 10 | opened pocket, fillet | 18 | opened island, opened pocket | 26 | simple hole |
| 11 | counterbore hole, closed pocket | 19 | simple slot | | |

a) Machining features in test case one

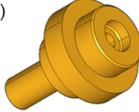

| No | List of machining feature | No | List of machining feature |
|---|---|---|---|
| 1 | simple hole, counterbore hole, closed pocket, opened pocket, fillet | 7 | simple hole, counterbore hole, simple slot, closed pocket, opened pocket |
| 2 | simple slot | 8 | simple hole, simple slot, closed pocket, opened pocket |
| 3 | simple hole, simple slot, opened pocket | 9 | simple hole, simple slot, opened pocket, fillet |
| 4 | simple slot | 10 | simple hole, simple slot, closed pocket, opened pocket |
| 5 | simple slot, opened pocket | 11 | opened pocket |
| 6 | closed pocket, opened pocket | 12 | floorless pocket, opened pocket |

b) Machining features in test case two

**Fig. 11. Machining feature recognition experiments for two test cases**

The proposed method correctly recognized all machining features included in test cases 1 and 2. However, the features included in the two CAD models in Fig. 12 were not correctly recognized. Fig. 12(a) is a floorless pocket with a steep base face, and Fig. 12(b) is a case where two base faces were merged into one face due to interference between the two features. We defined the value of the descriptor item $D_{ax\_interference}$ as 'True' to recognize the floorless pocket. This means that there is an intersection face when a ray is projected in the direction of the normal vector from the center of the base face. However, in the case of Fig. 12(a), there is no intersecting face with the ray of the base face because the base face has a relatively higher inclination. As a result, the features in Fig. 12(a) were misrecognized as opened pockets. In Fig. 12(b), the base faces of the two features were merged into one face due to interference between the simple slot and closed pocket. As a result, the similarity was calculated for the merged face. In this case, the two features must be recognized separately or as opened pockets, but they were misrecognized as simple slots.

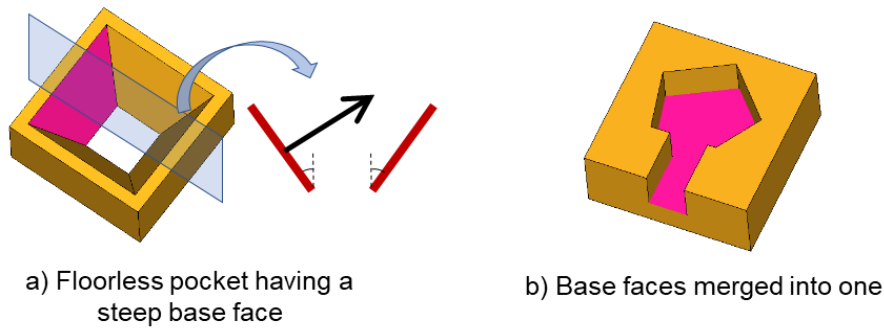

a) Floorless pocket having a steep base face

b) Base faces merged into one

**Fig. 12. Misrecognized machining features**

**5.2. Feature Recognition Performance Comparison with Hierarchical CADNet**

In this study, the recognition performance was compared with the latest artificial neural network to certify the superiority of the proposed method. Hierarchical CADNet(adj) was selected for performance comparison. Hierarchical CADNet uses a B-rep model as an input model. Due this characteristic, recognition performance can be compared using the same test cases. Hierarchical CADNet reported the highest accuracy and accuracy per face compared to other machining features recognition models such as FeatrueNet, MSVNet, and PointNet++. Therefore, the performance between the method proposed in this study and Hierarchical CADNet was compared.

To compare the performance between the method proposed in this study and Hierarchical CADNet, 40 models were selected from the MFCAD++ dataset[33] and defined as test case 3, as shown in Fig. 13. These models were used to verify the machining feature recognition performance. The models selected as test case 3 are 40 models listed at the top in the test model list file among the MFCAD++ dataset.

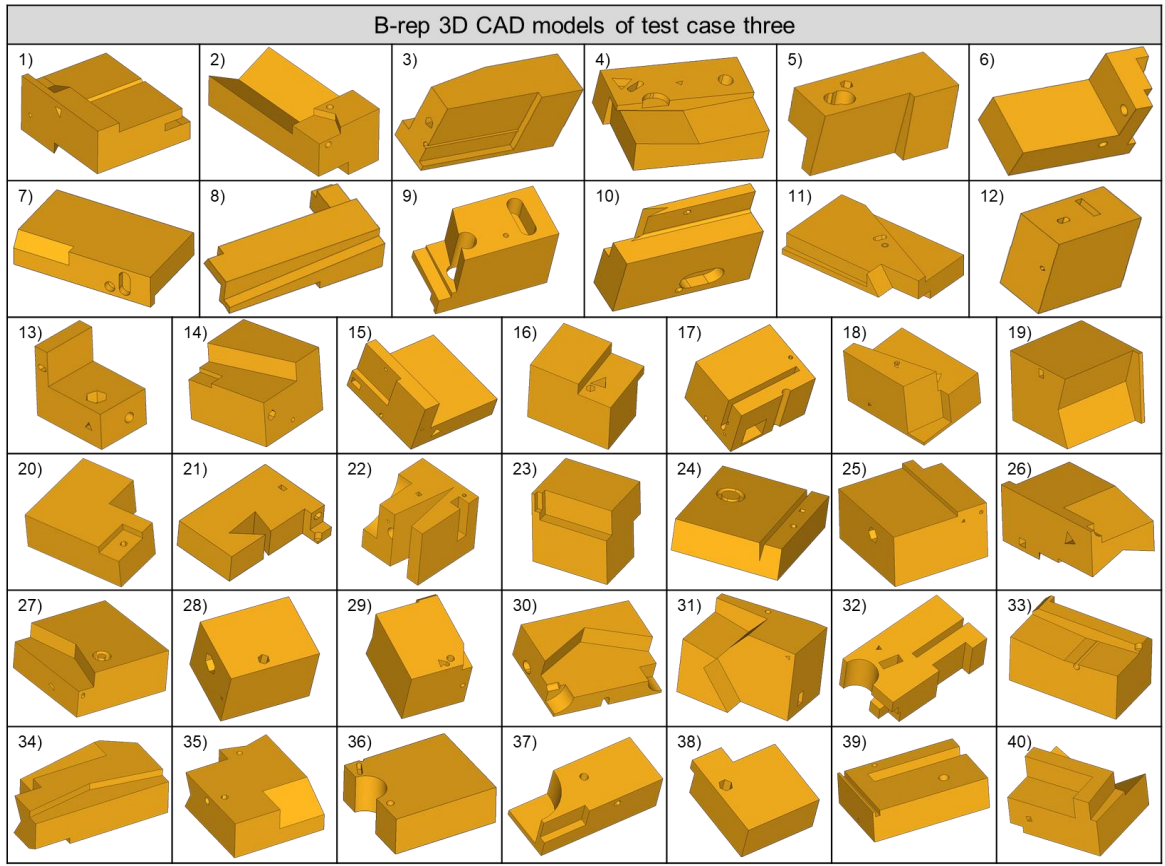

**Fig. 13. Machining features in test case 3**

Test case 4 was defined by excluding some models or changing some features of the models in test cases 1 and 2. Some models in test cases 1 and 2 have a rotational stock. However, Hierarchical CADNet does not support this type of stock. Therefore, models of this type were excluded from test cases 1 and 2. Also, test cases 1 and 2 have models that include machining features that Hierarchical CADNet does not support, such as opened island and counterbore hole. Therefore, if a 3D model has counterbore holes, these were replaced with simple holes. In addition, if a 3D model has an opened island, it was excluded from the test case. Test case 4[43], prepared as explained above, is shown in Fig. 14.

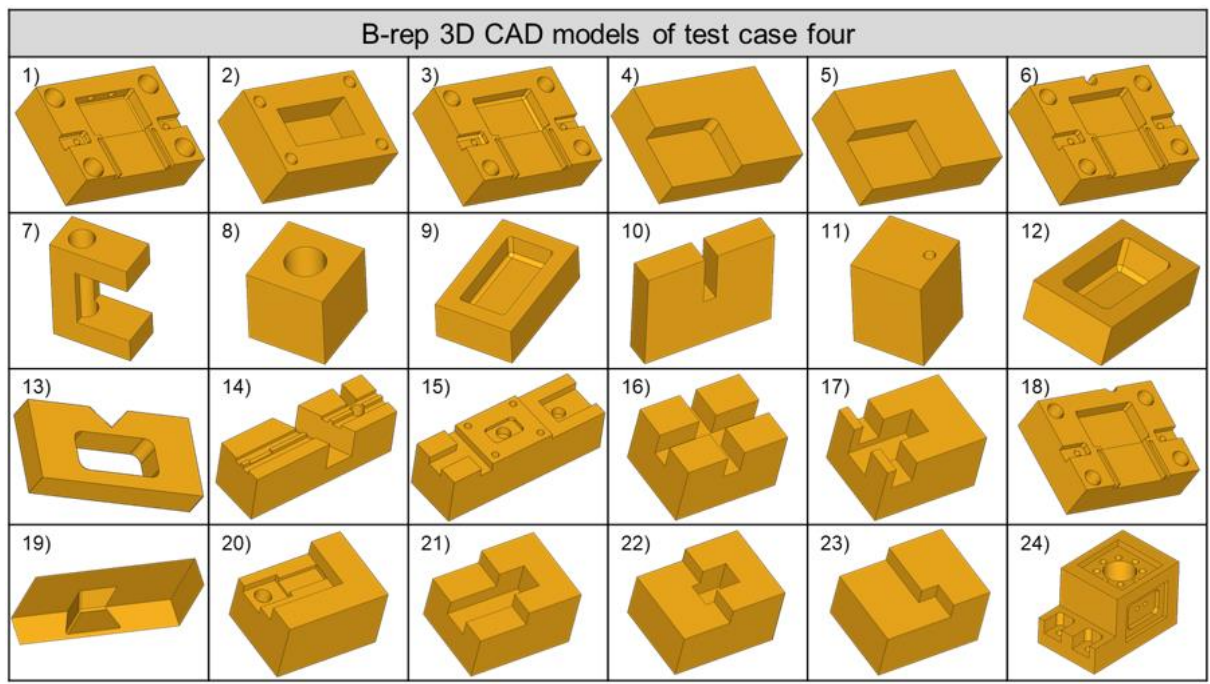

**Fig. 14. Machining features in test case 4**

The feature classification of this study and MFCAD++ are different. Accordingly, there are cases in which two classifications need to be differently determined, even if it is the same base face, such as slots and pockets. Therefore, new criteria for success in recognition of some machining features are necessary in the two cases: recognizing test case 3 by applying to the method proposed in this study and recognizing test case 4 by applying to hierarchical CADNet. Considering these cases, the machining feature classification of test cases 3 and 4 were reclassified into pocket type, slot type, O-ring type, hole type, chamfer type, and fillet type. The reclassification results are shown in Fig. 15.

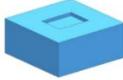

Fig. 15. Label classification of test cases 3 and 4

After reclassifying the machining feature, the criteria for recognition of slot/pocket, fillet, and O-ring are defined as follows when recognizing test cases 3 and 4 by the proposed method in this study and Hierarchical CADNet. First, the base faces A and B in Fig. 16(a) are classified into pocket and slot according to the method of classifying features in this study (whether the width of the feature matches the width of the tool). However, due to different classification criteria, two base faces are recognized as slot type when each base face is recognized using Hierarchical CADNet. Therefore, it was determined as a success if base face A in Fig. 16(a) is recognized as *Rectangular through slot*, a slot type label, when applying test case 4 to Hierarchical CADNet.

Next, the criteria for recognition success of fillet type and chamfer type were defined as follows. A face classified as fillet type, like base face C in Fig. 16(a), can be included in other features. Therefore, the corresponding face is recognized redundantly as a fillet type and other machining features in the method proposed in this study. However, Hierarchical CADNet recognizes the face as only one machining feature. Therefore, if the corresponding face is recognized as fillet type or feature type with fillet as an adjacent face, it was determined as a success when applying test cases 3 and 4 to Hierarchical CADNet. Also, if the corresponding face was recognized as fillet type and feature type with fillet as an adjacent face simultaneously, it was determined that recognition is successful when applying test cases 3 and 4 to the proposed method. The same method is applied even when the target machining feature to be recognized is a chamfer type.

Finally, the criterion for recognition success of the O-ring type was defined as follows. The machining feature D in Fig. 16(b) is classified as a O-ring in MFCAD++. However, the corresponding feature is classified as a composite feature of simple hole and closed island in the proposed method. Therefore, if the corresponding feature is recognized as a composite feature that consists of a simple hole and closed island, it was determined that recognition is successful when applying test case 3 to the method proposed in this study.

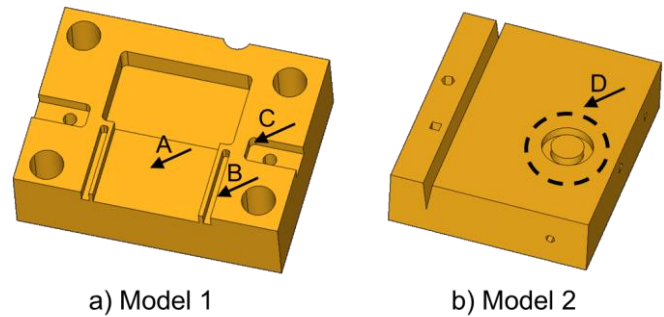

a) Model 1    b) Model 2

**Fig. 16. Faces that are differently classified between the proposed method and Hierarchical CADNet**

The performance of machining feature recognition was quantitatively calculated as follows. First, a multi-class confusion matrix was constructed based on the result of the recognition experiment targeting test models. Then, Precision, Recall, Accuracy, and F1 Score were calculated from the confusion matrix. Precision is the ratio of "the true faces that make the specific machining feature (*TP*)" to "all classified faces that make the specific machining feature (*TP+FP*)." Precision is defined as Eq. (7). Recall is the ratio of "the true faces that make the specific machining feature (*TP*)" to "the real faces that make the specific machining feature (*TP+FN*)." Recall is defined as Eq. (8). Accuracy is the ratio of "the correct face that make the specific machining feature (*TP+TN*)" and "all faces that make the 3D model (*TP+TN+FP+FN*)." Accuracy is defined as Eq. (9). Here, T, F, P, and N mean true, false, positive, and negative, respectively. F1 Score is a harmonic mean and has a characteristic that is less affected by the data bias, and thus, this metric is used. F1 Score is defined as Eq. (10).

$$Precision = TP/(TP + FP) \tag{7}$$
$$Recall = TP/(TP + FN) \tag{8}$$
$$Accuracy = (TP + TN)/(TP + TN + FP + FN) \tag{9}$$
$$F1\ Score = 2 \times (Precision \times Recall)/(Precision + Recall) \tag{10}$$

Through a performance comparison experiment between the proposed method and Hierarchical CADNet using test cases 3 and 4, it was determined whether the classification result for the faces that make up the feature was correct. First, a multi-class confusion matrix was created after comparing the value of real classifications and the value of prediction classifications from the recognition process. Then, Precision, Recall, Accuracy, and F1 Score were calculated based on this confusion matrix. As shown in Table 5, the proposed method outperforms Hierarchical CADNet in all evaluation metrics, such as Precision, Recall, Accuracy, and F1 Score.

**Table 5. Recognition performance of the proposed method and Hierarchical CADNet for test cases 3 and 4**

| The label | Data | Precision | Recall | Accuracy | F1 Score |
| --- | --- | --- | --- | --- | --- |
| Our method | Test case 3 | 0.9023 | 0.9676 | 0.9403 | 0.9338 |
|  | Test case 4 | 1.0000 | 1.0000 | 1.0000 | 1.0000 |
| Hierarchical CADNet | Test case 3 | 0.8751 | 0.8415 | 0.9301 | 0.8579 |
|  | Test case 4 | 0.6040 | 0.4099 | 0.4321 | 0.4884 |

Fig. 17 shows cases of recognition failure when the proposed method was applied to test case 3. In the first case, the width of the chamfer is larger than the width of the pocket, as shown in Fig. 17(a). We used a descriptor item *chamfer-machining* to classify the slot, pocket, and chamfer, meaning that the widths of slots or pockets should be larger than the widths of chamfers. However, the pocket in Fig. 17(a) could not be recognized correctly because the width of the chamfer was larger than the width of the pocket. In the next case, as shown in Fig. 17(b), the width of the pocket is smaller than the width of the slot. We used a descriptor item *face-machining* to classify the slot and pocket, meaning that the width of the pocket should be larger than the width of the slot. However, the pocket in Fig. 17(b) could not be recognized correctly because the width of the pocket was smaller than the width of the slot.

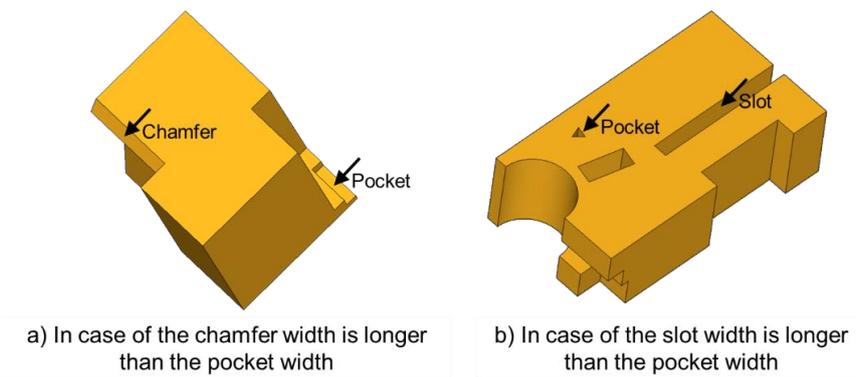

**Fig. 17. Failure cases when applying test case 3 to our method**

Fig. 18 shows the cases of recognition failure when Hierarchical CADNet was applied to test case 4. In the first case, a simple hole with two half cylindrical faces was misrecognized, as shown in Fig. 18(a). Hierarchical CADNet is unable to recognize holes with two half cylindrical faces because all holes in MFCAD++ dataset consist of only one cylindrical face. In the next case, the fillet in Fig.18(b) was misrecognized among the fillets.

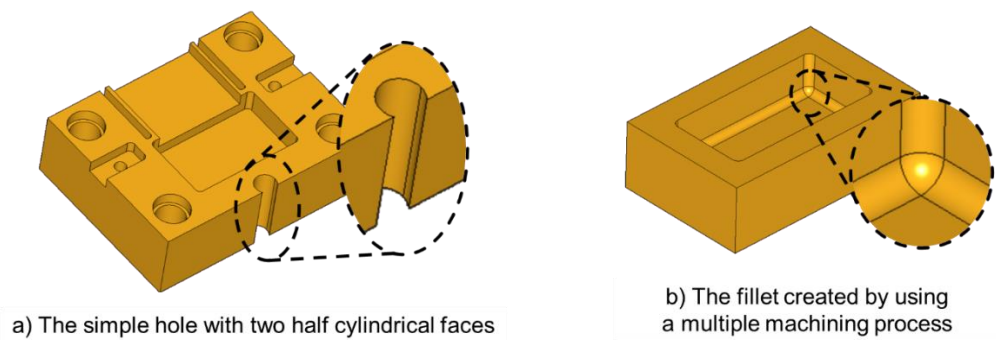

**Fig. 18. Failure cases when applying test case 4 to Hierarchical CADNet**

## 5.3. Comparison of Responsiveness to Stocks and Composite Features with Hierarchical CADNet

Additional experiments were performed to compare the responsiveness of the proposed method and Hierarchical CADNet to changes in the stocks or composite features. These experiments were conducted as follows.

First, recognition experiments were performed on rotational stocks, as shown in Fig. 19(a). The proposed method in this study recognized the rotational stock as well as the cuboid. However, Hierarchical CADNet could not recognize the rotational stock as well as machining features in rotational stock. Through these results, we certified that the deep learning-based method must undergo additional learning when there is stock shape change.

When performing a recognition experiment on a counterbore hole consisting of a combination of two simple holes, as shown in Fig. 19(b), the proposed method recognized the counterbore hole as two simple holes if there was no descriptor for the counterbore hole. However, Hierarchical CADNet misrecognized the counterbore hole as a completely different machining feature. Through these results, we certified that a new label must be additionally created according to the feature when a combination of features is recognized.

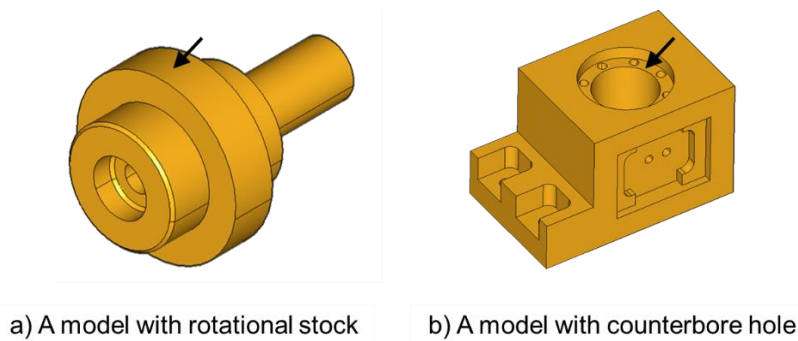

Fig. 19. Additional test models used to compare the proposed method and Hierarchical CADNet

## 6. Conclusions

In this study, we proposed a machining feature recognition method that compares the similarity of feature descriptors using the concept of range constraints, including *minimum*, *maximum*, and *equal* constraints. The minimum and maximum constraints refer to the lower and upper bounds of the values that each descriptor item can have, respectively. Furthermore, the *equal* constraint refers to the value that each descriptor item must have. After implementing the prototype system, feature recognition tests were conducted for two test cases. The improved descriptor supports the recognition of 16 types of machining features. The results showed that the recognition performance improved remarkably; all machining features included in the two test cases were successfully recognized. In addition, we confirmed that the recognition performance of the proposed method in this study is higher than the latest artificial neural network by comparing the two methods. We also confirmed that the proposed method has good responsiveness to 3D models including unused stock or features in the experiment.

In the future, we plan to expand the types of features covered to recognize 3D shapes of functional parts expressed by the combination of multiple features and general machining features. Furthermore, the descriptors proposed in this study will be revised and improved to solve the misrecognition problem described in Fig. 12. In addition, we will conduct studies to recognize machining features by applying the proposed descriptors to artificial deep neural networks such as convolutional neural networks and recurrent neural networks.

**Ethical Approval**

This manuscript has not been published or presented elsewhere in part or in entirety and is not under consideration by another journal.

**Consent to Participate**

Not applicable

**Consent to Publish**

Not applicable

**Authors Contributions**

**Seungeun Lim**: Methodology, Data Curation, Software, Writing - Original Draft. **Changmo Yeo**: Data Curation, Software, Writing - Original Draft. **Fazhi He**: Resources, Writing – Review & Editing. **Jinwon Lee**: Methodology, Validation, Investigation, Writing - Review & Editing **Duhwan Mun**: Supervision, Conceptualization, Methodology, Writing - Review & Editing, Funding.

**Acknowledgements**

This research was supported by the Basic Science Research Program [No. NRF-2022R1A2C2005879] through the National Research Foundation of Korea (NRF) funded by the Korean government (MSIT), by the Carbon Reduction Model Linked Digital Engineering Design Technology Development Program [No. RS-2022-00143813] funded by the Korean government (MOTIE), and by Institute of Information & communications Technology Planning & evaluation (IITP) grant funded by the Korea government (MSIT) [No.2022-0-00969].

**Competing Interests**

The authors declare no potential conflicts of interest with respect to the research, authorship, and publication of this article.

**Availability of data and materials**

Not applicable